# The role of sex separation in neutral speciation

Elizabeth M. Baptestini[1], Marcus A.M. de Aguiar[1,2] ,Yaneer Bar-Yam[2]

[1] Instituto de Física Gleb Wataghin

Universidade Estadual de Campinas,

Campinas, SP, 13083-970, Brazil

E-mail: aguiar@ifi.unicamp.br

[2] New England Complex Systems Institute,

Cambridge, Massachusetts 02142, USA

**Abstract -** Neutral speciation mechanisms based on isolation by distance and sexual selection, termed topopatric, have recently been shown to describe the observed patterns of abundance distributions and species-area relationships. Previous works have considered this type of process only in the context of hermaphrodic populations. In this work we extend a hermaphroditic model of topopatric speciation to populations where individuals are explicitly separated into males and females. We show that for a particular carrying capacity speciation occurs under similar conditions, but the number of species generated decreases as compared to the hermaphroditic case. Evolution results in fewer species having more abundant populations.

## I - Introduction

Natural selection is a key process in the adaptation of species to changes in the environment and to changes in other species. Whether it plays an important role in shaping the observed patterns of biodiversity, however, has been questioned. Neutral theories, based on drift and statistical fluctuations in populations size alone, have been very successful in reproducing the observed abundance distributions, which exhibit remarkable universal features (Hubbell 2011, Kopp 2010, Etienne et al. 2011, Rosindell et al 2011).

Speciation is the ultimate driver of biodiversity. This process, however, has received relatively little attention in the context of neutral models. The neutral theory of biogeography developed initially by MacArthur and Wilson (1967), Hubbell (2001) and others (Kopp 2010; TerSteege 2010; O'Dwyer et al. 2010) included speciation as random point mutations, without specifying an underlying mechanism. The assumption of random speciation events may be reasonable for describing island biogeography, where `speciation' represents the arrival of new species from the continent. However, the point mutation model is a more radically simplifying assumption otherwise. Requirements for multiple individuals in a viable sexually reproducing population and the role of subpopulation divergence should be considered. Hence, the process of speciation merits a discussion of its own in the neutral framework (Banavar et al. 2009; Rosindell et al. 2010, Etienne et al. 2011).

Among the many types of speciation processes, allopatry is considered to be the dominant form (Mayr 1988). Allopatric speciation happens when the genetic flow between groups of individuals is blocked by geographic barriers. These isolated groups evolve independently, either by selection or drift, eventually acquiring incompatibilities leading to reproductive isolation. In this context, the evidence for the role of natural and sexual selection in promoting reproductive isolation has been observed both in laboratory experiments and in nature. Neutral divergence, due to drift alone, has also been observed in plants and mammals, although in fewer cases (Coyne & Orr 2004).

Sympatric speciation, on the other hand, is triggered by ecological interactions taking place in a single spatial domain and even in the same niche (Rosenzweig 1997). The key driver is the coupling between ecological and mating traits, which may lead to disruptive selection and ultimately to speciation (Dieckmann et al. 1999; Leimar et al. 2008; Baptestini et al. 2009; Pinho & Hey 2010).

A neutral theory of speciation relying on isolation by distance (Wright 1943), without geographic barriers or ecological interactions, has been recently demonstrated (de Aguiar et al. 2009). The mechanism, termed topopatry, was shown to describe the universal features observed in abundance distributions and species-area relationships. In this context, sexual selection driven by spatial and genetic distances with neutral and independent genes is sufficient to promote speciation. The demonstration that speciation can happen *even* in homogeneous environments also suggests that speciation can be accelerated by the presence of partial barriers, selection and gene interactions. Interesting examples are provided by ring species, where geography plays a crucial role in physically shaping the ring but without blocking dispersal or gene flow along the ring (Irwin et al. 2001, 2005; Ashlock et al. 2010). Perhaps the most important feature of topopatric speciation is the possibility that all genetic mechanisms leading to reproductive isolation in allopatric processes may act without the need for allopatry itself, i.e., without an initial period of geographical separation.

Allopatry and sympatry can be viewed as extremes of a continuum of speciation modes (Fitzpatrick et al. 2009). Topopatry, based on isolation by distance, is a new example of an intermediate case.

The topopatric model described in de Aguiar et al (2009) relied on several limiting assumptions that should be relaxed for comparison with real ecologies. In particular, the individuals were considered haploid and hermaphroditic. In this paper we consider males and females explicitly and study the effects of sex separation in the process of speciation.

The role of sex separation in evolution has been shown to either favor or hinder speciation, depending on the mechanisms driving it. An important case is that of diploid organisms where Dobzhansky-Muller incompatibilities related to sex chromosomes exist (Orr 1997, Orr & Presgraves 2000, Turelli & Orr 2000, Kondrashov & Kondrashov 2001, Coyne & Orr 2004, Haerty & Singh 2006). According to this perspective, sexual differentiation facilitates speciation through unviable hybrids according to Haldane’s rule. On the other hand, the separation of individuals into males and females may give rise to sexual dimorphism, where significant phenotypic differences between the two sexes develop. Recent studies have shown that adaptive speciation and ecological sexual dimorphism may compete as outcomes of assortative mating, restricting the likelihood of speciation (Bolinick et at. 2003, Parker & Partridge 1998).

In this work we consider the primary effects of sex separation on topopatric speciation. Ecological traits, direct competition or epistatic effects between sexual and asexual chromosomes are not considered. Our treatment, although simplified, has the advantage of isolating the effects of sex separation in the process of reproduction. We find that for a particular carrying capacity speciation occurs for similar conditions as described by model parameters, but the number of species decreases by a factor of about two as compared to the hermaphroditic model. Evolution in this case results in less but more abundant and stable species.

The paper is organized as follows: in Section II we provide a detailed description of the model and of our working definition of species. In Section III we discuss some theoretical predictions based on results obtained for hermaphroditic populations and in Section IV we present the results of numerical simulations. Finally, in Section V we present our discussions and conclusions.

## II - The Model

We use an agent-based model to simulate the neutral evolution of spatially distributed populations. The current approach differs from that used previously (de Aguiar et al. 2009) in that we distinguish male and female individuals and restrict mating accordingly. In this section we present a detailed description of the model.

### *The physical and genetic spaces*

We consider an initial population of N genetically identical individuals randomly distributed over a homogeneous environment, represented by a rectangular geographical domain subdivided into L x L regions. We use periodic boundary conditions so that there are no boundaries or corners. Multiple individuals can exist at the same site but typically do not, since the density is low. The number of individuals is held fixed throughout the simulation, modeling an underlying fixed ecological capacity.

Each individual in the population is located at a position (x,y) in the physical space, with $1 \leq x,y \leq L$ and has a haploid genome, as illustrated in fig. 1. The genome consists of B+1 independent biallelic genes, which are labeled 0 or 1. The k-th gene of the i-th individual is denoted by $\sigma_k^i$ and the genome by the binary string

$$g^i = ( \sigma_1^i, \sigma_2^i, \ldots, \sigma_B^i; \sigma_{B+1}^i ). \quad (1)$$

The last gene determines the gender of the individual, $\sigma_{B+1}^{i} = 1$ for males and $\sigma_{B+1}^{i} = 0$ for females. At the beginning of the simulation all individuals have identical genomes with $\sigma_k^{i} = 0$ for k=1,…,B. The value of $\sigma_{B+1}^{i}$ is assigned randomly with equal probability of male and female.

The time evolution of the population is governed by sexual reproduction, mutation and recombination and also by dispersal of the offspring. The key ingredient of the model is the introduction of assortative mating based on two critical mating distances (de Aguiar et al. 2009): one in physical space and one in genetic space. In physical space, an individual can mate only with others of opposite sex living in a certain neighborhood of its location determined by the spatial mating distance S. This type of spatial mating restriction was considered by Sewall Wright (1940, 1943) and Kimura & Weiss (1964) and may lead to significant genetic differences between geographically distant individuals of the same species. Striking evidence of this mechanism of "isolation by distance," is provided by ring species (Irwin et al. 2001, 2005).

Spatial proximity is a necessary but not sufficient condition for mating. We also assume that individuals that are too different genetically will not be able to mate successfully. Among the many reasons for this are structural differences in the sex organs, failure of the sperm to reach or fuse with the egg or the failure to elicit mating behavior (Coyne & Orr 2004). As shown previously (de Aguiar et al. 2009) the genetic restriction on mating alone does not lead to speciation, but it keeps different species (after they have been formed) genetically isolated from each other. To impose genetic proximity on mating organisms (Gavrilets 2004; Higgs et al. 1991, 1992) we restrict the number of distinct genes to be no more than the genetic mating distance G. The genetic distance between individuals i and j is measured by the Hamming distance

$$d(i,j) = \sum_{k=1}^{B} \left| \sigma_k^i - \sigma_k^j \right| \qquad (2)$$

and mating is possible if $d(i,j) \leq G$.

## *Time evolution*

The evolution of each generation is divided into N *time steps*, in which a single individual reproduces. After N such steps the entire population has been replaced. We start with the ith individual which attempts to reproduce and is successful with a probability (1-Q). We identify all individuals of opposite sex in its spatial mating neighborhood, specified by the distance S, and whose genetic distance is less or equal to G. From this list one is randomly chosen as a mate, say, individual j.

The genome of the offspring is obtained by a single recombination of $g^i$ and $g^j$: a random position k in the parent's genomes is chosen to cross-over and two new genomes, $g^a$ and $g^b$, are produced:

$$g^a = ( \sigma_1^{\,i}, \sigma_2^{\,i}, \ldots, \sigma_k^{\,i}, \sigma_{k+1}^{\,j}, \ldots \sigma_B^{\,j}; \sigma_{B+1}^{\,j} )$$

$$g^b = ( \sigma_1^{\,j}, \sigma_2^{\,j}, \ldots, \sigma_k^{\,j}, \sigma_{k+1}^{\,i}, \ldots \sigma_B^{\,i}; \sigma_{B+1}^{\,i} ) \quad (3)$$

One of these is taken randomly as the offspring's genome, which is further subjected to mutations, at a rate μ per gene. The process is illustrated in fig.2.

The offspring is placed at position $(x^i,y^i)$ with probability 1-D and, with probability D, at random within a small region of radius r around $(x^i,y^i)$. D is the dispersal rate and r the dispersal range. After reproduction the originating parent expires and the label i is assigned to the offspring. A generation is comprised by the reproduction of all N individuals in sequence, from i=1 to i=N. Note that this does not imply any spatial ordering in reproduction, since the individuals are randomly placed at the beginning. However, the generations are partially overlapping, since a newly born offspring can be chosen as mate partner of another individual during the same `mating season'.

Reproduction of the ith individual is, however, only successful with probability 1-Q. With probability Q the individual dies without leaving a descendant. In this case another individual, chosen at random within the spatial neighborhood of radius S, reproduces instead of the original individual. The offspring generated is placed in the position of the original individual or in its neighborhood according to D and r. On average, two offspring are born for each parent. If Q=0 each individual has at least one offspring and also has a probability of being selected by neighbors as a mate for one of their offspring. If $Q \neq 0$, some have none and the distribution of numbers of offspring includes those with additional offspring to offset them.

During the selection process restricted by spatial and genetic proximity, it might happen that the number of mates available to the reproducing individual is very small, possibly zero, preventing it from finding a mate. To avoid this situation we introduce the parameter P, representing the minimum number of *potential mates*. Given S and G, if the number of mates available to the individual is smaller than P, we relax the spatial constraint by increasing $S \rightarrow S + 1$ for the present mating season only, i.e. the individual increases the search area in order to have more choices. If the number of available mates is still smaller than P, the process is repeated until S increases up to 10 units. If the number of mates is still smaller than P we pick another organism in the original neighborhood to reproduce. In the paper by de Aguiar et al. (2009) the genetic constraint was also relaxed simultaneously with S, i.e. $S \rightarrow S + 1$ *and* $G \rightarrow G + 1$. Here we let only S change and keep the genetic restriction fixed at all times.

## *Species*

Many definitions of species have been proposed that work well in specific groups of organisms but fail, or are impractical, in others. The most used of these definitions is perhaps Ernst Mayr's

Biological Species Concept (BSC) (Mayr 1995), based on the interbreeding ability of the individuals in a group. Another concept is that of genetic cohesion due to Mallet (1995), termed Genotypic Cluster Species Concept (GCSC), according to which a species is a genetically distinguishable group of organisms that has no (or few) intermediates when in contact with other such groups. A similar definition, is the Cohesion Species Concept due to Templeton (1989).

For our purposes a species is a group of individuals related by potential gene flow, which need not be possible in a single generation. As an example, consider three individuals A, B and C such that $d(A,B) < G$, $d(B,C) < G$ but $d(A,C) > G$. A mutation occurring in A can be transmitted to the offspring of A and B that can, in turn, pass the mutation on when mating with C or its offspring. This situation is common in ring species (Irwin et al. 2001, 2005), and we also find it occurs in our simulations. In the case of a ring species, the appearance of an advantageous mutation on a few individuals might spread over entire ring, due to its genetic cohesion. This, however, might take multiple generations. According to the BSC A and C should belong to different species. However, A, B and C belong to the same species in the GCSC, since in genetic space, the individuals form a cluster that is cohesive and is separtated by more than G from all organisms not in the cluster. Thus our definition is similar to, if not exactly the same as, GCSC.

In order to classify the individuals in the population into species, the following algorithm is applied: we start with individual number 1 (which is arbitrary) and assign it to the first species, Species-1. We collect all others such that $d(1,i) \leq G$ and assigned them to Species-1. For each of the individuals i just added to Species-1 we check if $d(j,i) \leq G$ for all unassigned individuals. The individuals satisfying this condition are also assigned to Species-1. For these new individuals j we check again if $d(k,j) \leq G$ for all unassigned k. The individuals satisfying this condition are also added to Species-1 and so on. When no more individuals are added, Species-1 in completed. It is a cohesive group and genetically isolated from the unassigned individuals. If there are no unassigned individuals, there is only one species. Otherwise, we take one unassigned individual and assign it to the second species, Species-2, repeating the process. It is straightforward to prove that the species obtained in this way are independent of which individuals are chosen. Note that the only criterion used to define species is the genetic mating distance G. No information about the spatial location of the individuals is taken into account.

## III - Theoretical Results

A population whose individuals are genetically identical at time zero, develop differences through mutation, which occurs at the rate $\mu$. These differences, however, are constrained by sexual reproduction, which tends to contract the genetic spreading caused by mutations. The balance between these two opposing forces results in the natural diversity of the population. When spatial and genetic selection are present, characterized by S and G, for some values of the parameters the outcome of evolution is the spontaneous breakup of the population into multiple species. The number of species formed depends on the parameters of the model and can be estimated by equation (8) below. This type of pattern formation, coupling the genetic and physical spaces, has been observed before in simpler systems (Sayama et al. 2002) and also in speciation models (Hoelzer et al. 2008).

The hermaphroditic model can be mapped into an influence dynamical process on networks, from which a number of analytic results can be extracted (de Aguiar & Bar-Yam, 2011). An example is the determination of a critical line in the G versus S plane below which speciation occurs. It is given by

$$G_c(S) = \frac{B/2}{1 + \mathrm{P}_r^{-1}} \tag{4}$$

where

$$P_r = \exp\left( -\frac{\pi^2 \rho_0^2 (S - S_{\min})^4}{B^2 \mu^2 \gamma^4} \right), \tag{5}$$

$\rho_0 = N/L^2$ is the average population density, $S_{\min} = \sqrt{P/\pi\rho_0}$ is the size of a neighborhood containing P individuals and $\gamma$ is a parameter obtained by fitting to simulations.

Particularly important for the present discussion is the calculation of $N_s$, the number of species that arise from speciation. It can be written as $N_s = N/N_i$ where $N_i$ is the average number of individuals in a species, which in turn can be written as

$$N_i = \pi \rho_0 R^2. \tag{7}$$

where R is a measure of the species spatial extent. For a hermaphroditic population (de Aguiar & Bar-Yam 2011)

$$N_i = \frac{G}{2\mu(B_{ef} - 2G)}$$

where $B_{ef}$ is the effective number of genes, given by $B_{ef} = 2G - (B - 2G)P_r$. For panmictic populations, $B_{ef}$ approaches 2G. reflecting the strong restriction imposed by the condition of genetic proximity in mating. As mating becomes spatially constrained by decreasing S, the effective role of mutations increases and so does $B_{ef}$. Speciation occurs when the average genetic distance between the individuals increases from G (in the panmictic case) to approximately 2G. Combining these equations we obtain

$$N_s = \frac{2\mu\rho_0 L^2}{G}(B - 2G)\ P_r \tag{8}$$

Assuming that $P_r \approx 1$, the exponential in equation (5) can be expanded to first order in its argument. Using $N = N_s(\pi \rho_0 R^2)$ we obtain

$$R \simeq \sqrt{\frac{G}{2\pi\mu\rho_0(B-2G)}\left[1+\frac{\pi^2\rho_o^2(S-S_{\min})^4}{B^2\mu^2\gamma^4}\right]} \quad (9)$$

for the average radius of a species. A similar expression was obtained in (de Aguiar et al 2009) using simulation results.

Equations (4), (8) and (9) fit well the numerical data in the hermaphroditic case. In the model with sex separation these estimates also work reasonably well if the equations are properly adapted. In the hermaphroditic case the population can be seen as a network where each individual is a node and links are established between potential mates (spatially close and genetically compatible). In the case of sex separation, the nature of the network changes considerably: the nodes represented by females don't link among themselves, but only with nodes representing the males, and vice-versa. Networks made up of two disjoint sets of nodes, such that nodes in one set connect only to nodes in the other set are called *bipartite* networks. It is useful to define a female network, where two individuals are connected if they can mate with a common male, and similarly for a male network. For each of these networks, having approximately half of the population, the average density is $\rho_0/2$. Also, if the spatial restriction between males and females is *S*, the maximum separation between individuals of the same sex in a species is *2S*. Moreover, gene flow in the female network is at least a two step process, since it takes the mating of a first female to a male and, if the offspring is male, its mating with a second female to combine their genes. Therefore, the time scale of crossover events slows down by a factor of two. Still, the analytic discussion continues to apply to the female or male network so the form of the analytic expressions is valid. We find from simulations described below that equations (4), (8) and (9) can indeed be applied to the sexual model if we change $\rho_0$ to $\rho_0/2$, $S_{min}$ to $\sqrt{2}S_{\min}$ and *S to 2S.* In order to obtain good qualitative fits we also need to change $\gamma$ into $2\gamma$.

## IV - Results of Simulations

The purpose of the present numerical simulations is to study the patterns of abundances resulting from the explicit introduction of males and females and compare them with those obtained with the hermaphroditic model. Because both models involve sexual reproduction, we refer to former as the *sex separated* model.

Since the number of model parameters is quite large we will keep many of them constant throughout the simulations. Variations in these parameters do not alter significantly the qualitative results. The fixed parameters are: mutation rate $\mu$=0.001; length of genome B = 125;

diffusion rate D = 0; minimum number of potential partners P = 5; probability of no reproduction Q = 0.3. In most cases we will also use S=5, G=20, N=8000 and L=256. In all figures time is measured in number of generations. Because there is no difference in the fitness assigned to males and females, the sex ratio is nearly constant across generations, fluctuating around N/2, as shown in figure 3.

Figure 4(a) shows the number of individuals between r and r + 1 as measured from the geographic center of a species and averaged over all species (squares) compared with a fit by $r \exp(-r^2 / R^2)$. Figure 4(b) shows the genetic distance between individuals of a species as measured from a reference individual situated closest to the geographic center of the species. This shows clearly the strong correlation between spatial distance and genetic distance. It also shows that the central individual can mate with any member of the population as far as the genetic constraint is concerned (i.e. the average asymptotic genetic distance is 14.3 which is less than G=20), just as is necessary for the species definition according to BSC. In many cases, however, individuals at opposite spatial extremes of a species may have genetic distance larger than G and would not be able to mate even if brought spatially close to each other, a feature also observed in the hermaphroditic model.

One of the strengths of the Topopatric speciation model is its ability to replicate observed distributions of abundance. Typical abundance distributions are well fit by the lognormal function with excess rare species (May 1975; Sugihara 1980). In figure 5(a) we compare the results of simulations (black squares) with a lognormal (solid line) taking into account all species formed (sampled area equal to the total area of the lattice, 512 x 512). Figure 5(b) shows the distribution for sampling area corresponding to only one eighth of the total area available (128 x 128), displaying a clear excess of rare species as compared to the lognormal distribution. The distribution obtained for even smaller sample areas converge to a Fisher curve, fig. 5(c) (64 x 64). Fig. 5(d) shows the abundance-rank plot corresponding to panels (a)-(c), displaying the typical S-shaped curve for large sampling areas.

Figure 6 shows a comparison of the Species Area Relationship (SAR) (Preston 1960; May 1975) for the hermaphroditic model, 6(a), and the model with sex separation, 6(b). Both display a triphasic pattern (Rosenzweig 1995; Tjorve 2003) of the form $A^z$ (Arrhenius 1921), with larger exponents at both the smallest and largest area regimes.

Figure 7(a) compares the number of species generated after 1500 generations (when equilibrium has already been reached) as a function of the total number of individuals in the population. The higher curve (black squares) represents the hermaphroditic model, whereas the lower curve (red circles) displays the result of the sex separated model. In both cases speciation is not possible at very low densities, since no (or very few) mating partners can be found in the search area delimited by S. More importantly, speciation is also inhibited at high densities, since statistical fluctuations, which play an important role in the process, decrease. The separation of individuals into males and females reduces the number of species formed and also prevents speciation for smaller populations. Figure 7(b) shows how the number of species changes with time. For the sex separated case equilibration takes about twice as long and the final number of species is halved, as expected from the theoretical arguments.

This effect can also be seen in figures 8(a) and 8(b), which shows the number of species formed in terms of the parameters S and G responsible for the sexual selection. Speciation occurs for both models for about the same range of S and G, for large S and G, although fewer species are formed in the sex separated case. The thin line shows the critical speciation curve according to the prediction of the hermaphroditic model, equation (4). The solid thick curve shows the same theoretical prediction with the changes $\rho_0$ into $\rho_0/2$, S into 2S and $\gamma$ into $2\gamma$. The shape of the level curves is similar, but in the sex separated case speciation is more severely hindered at low values of G, an effect already noted in de Aguiar et al. (2009) in the hermaphroditic model but is not as evident for the present values of parameters.

## V - Discussion

Speciation can be triggered by several processes, including geographic isolation, competition for resources, and genetic drift, among others. If the genes involved in speciation do not affect the fitness of the individuals, the speciation is termed “neutral”. The idea of neutral evolution, where the role of natural selection is secondary, has been challenged by many. Hubbell (Hubbell 2001), however, demonstrated that realistic patterns of abundance distribution can be obtained within a neutral theory of biogeography in which species originate randomly (Banavar & Maritan 2009, Kopp 2010, Ter Steege 2010, Etienne & Haegeman 2011, Rosindell et al. 2011).

Numerical simulations with hermaphroditic populations (de Aguiar et al 2009) have shown that similar patterns of diversity emerge in explicitly genetic neutral models if reproduction is constrained by spatial and genetic proximity between individuals, and quantitative agreement between observed and simulated diversity was also obtained using this model. Many of the results observed in these numerical simulations where recently derived analytically by mapping the genetic evolution into an influence dynamical process on networks (de Aguiar & BarYam 2011). In this paper we extended the hermaphroditic neutral model to describe speciation in populations with explicit sex separation.

The distinction between males and females changes considerably the genetic flow in a population. Unlike hermaphroditic species, where gene flow is allowed between any two members, here the individuals are divided into two separate groups with no direct gene flow within the groups, only between them. The mathematical description of this process, even for a panmictic population, is very different from the hermaphroditic case. Equations similar to the Moran model (Moran 1958, Cannings 1974, Ewens 1979, Gillespie 2004) can be written down, but explicit solutions are not available, except for zero mutation. In this trivial case it is possible to show that the equilibrium population is composed of identical individuals and the number of males and females follow a binomial distribution. Here we found that the analytic solutions obtained for the hermaphroditic model also approximately apply to the sex separated case if some re-scaling of the parameters is done.

For simulations, the extension of the model from hermaphroditic to sex separated individuals is constructed by adding an extra `gene' that specifies the sex and by restricting mating to individuals of the opposite sex. The main difference between the two models is that sexually separated individuals have, on average, half the number of potential mates than a hermaphroditic group with the same density of individuals. This might suggest that gene flow is more restricted when the two sexes are considered explicitly and that speciation should occur more easily. This, however, is not the case. The reason is that a population consisting of half males and half females is very different from two independent hermaphroditic populations with half the density. Sexual reproduction has a strong effect on the genetic proximity of offspring and is capable of keeping the population united.

We have shown that all basic features of the hermaphroditic model are preserved in the modified version, with few but important changes. Aside a small reduction in the parametric region where speciation occurs (fig.8), the most striking feature of the sex-separated model is the decrease in the number of species formed or, conversely, an increase in the average abundance of individuals per species, as shown in fig. 7. This is an important characterization of the model, which implies there are smaller extinction and re-speciation rates and, therefore, more stable species.

The results of our simulations cannot be directly applied to treat Dobzhansky-Muller type genetic incompatibilities or the development of sexual dimorphism, since the sex chromosomes are not considered explicitly in the model. However, extensions in this direction are possible.

## Acknowledgments

This work was supported by CNPq and FAPESP.

# Figure Captions

Figure 1. Snapshot of agent population on a square lattice. Individuals are represented by a square at its lattice location. The schematic shows the genome of a female for B = 11 and its mating neighborhood of radius S.

Figure 2. The process of reproduction with recombination and mutation. The top left show the genomes of the parents and the crossover point. After recombination one of the two resulting genomes is chosen for the offspring, which is then subjected to mutation, so that each gene can flip from 1 to 0 or vice versa with probability $\mu$.

Figure 3. Number of males and females as a function of time for S = 5, G = 20, N = 8,000 and L = 256.

Figure 4. (a) Spatial distribution of individuals, n(r), for the sex separated model averaged over all species. The simulation is shown as black squares and the Gaussian fit by the solid line (R2 = 0.989); (b) Correlation between genetic ($d_G$) and spatial ($d_S$) distances between individuals within a species averaged over all species. In both cases N = 8,000, L = 256, G = 20 and S = 5.

Figure 5. Species abundance distribution for 32,000 individuals, for S = 5 and G = 20. Simulations are shown as black squares and lognormal fits as solid lines. (a) for 512 x 512 lattice (R2 = 0.991); (b) for 128 x 128 sub-lattices (R2 = 0.856); (c) for 64 x 64 sub-lattices (R2 = 0.852). Panel (d) shows the abundance versus species rank for the same cases. Data was generated running 10 simulations for 1,500 generations and collecting all data.

Figure 6. The classical triphasic species area curve (SAR) for (a) hermaphroditic model; (b) the sex separated model. Parameters of the simulation are S = 5, G = 20, N = 32,000 and t=1500 for a 256 x 256 lattice.

Figure 7. (a) Number of species formed as function of the number of individuals in the population for the hermaphroditic (black squares) and sex separated (red circles) models. Parameters: S = 5, G = 20, L = 256, t=1500. (b) Time evolution of the number of species in the population for N=8,000.

Figure 8. Contour plot of the number of species formed as function of S and G for (a) the hermaphroditic and (b) the sex separated models. Parameters: N = 8000, L = 256, t = 4000. The thin solid line shows the critical curve according to equation (4) and the thick line in panel (b) shows the same curve with the re-scalings $\rho_0 \rightarrow \rho_0/2$, $S \rightarrow 2S$ and $\gamma \rightarrow 2\gamma$.

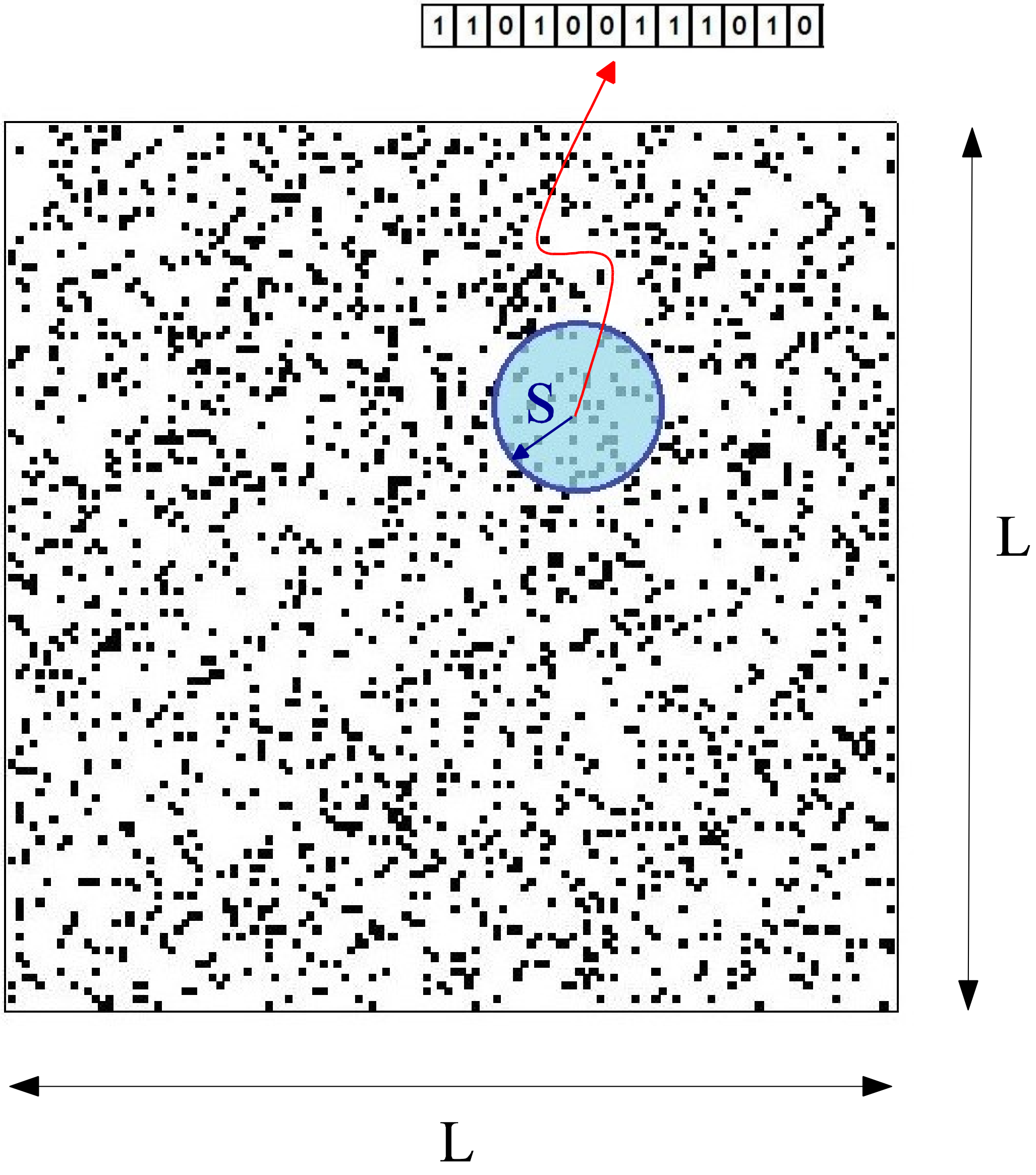

Figure 1

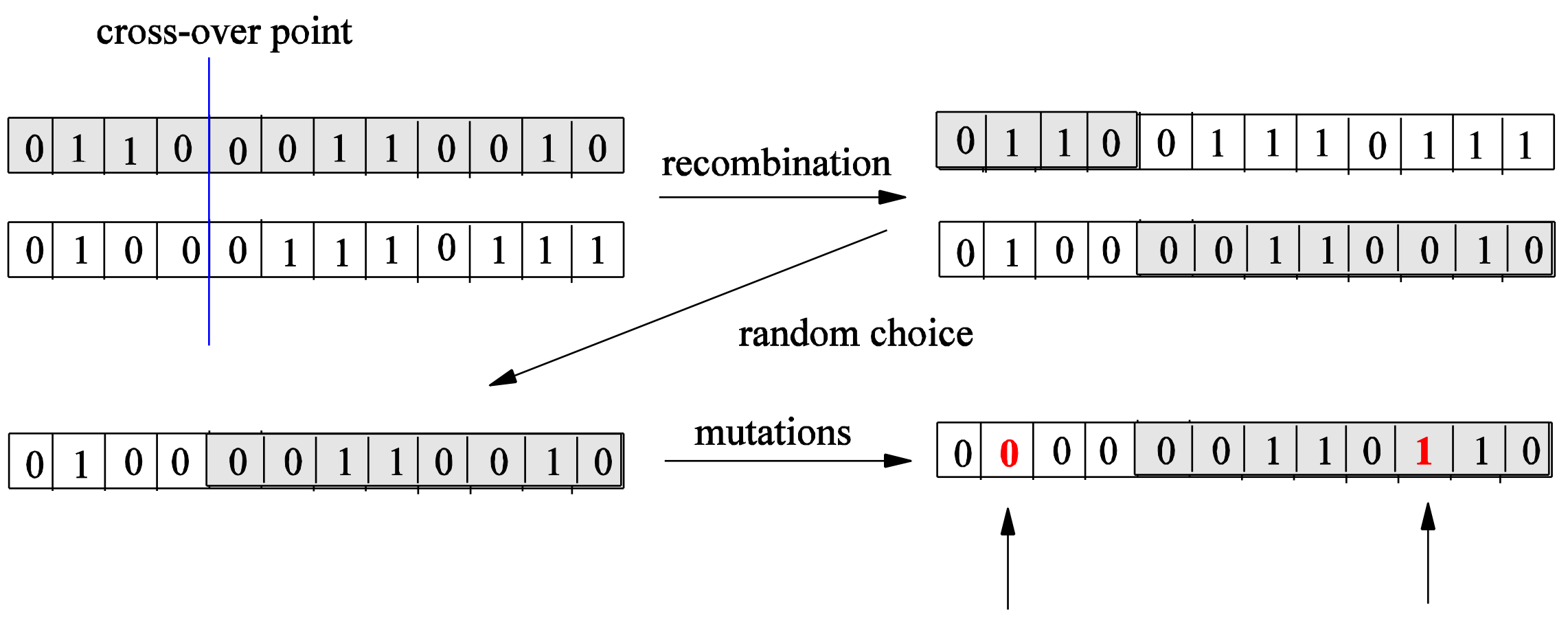

Figure 2

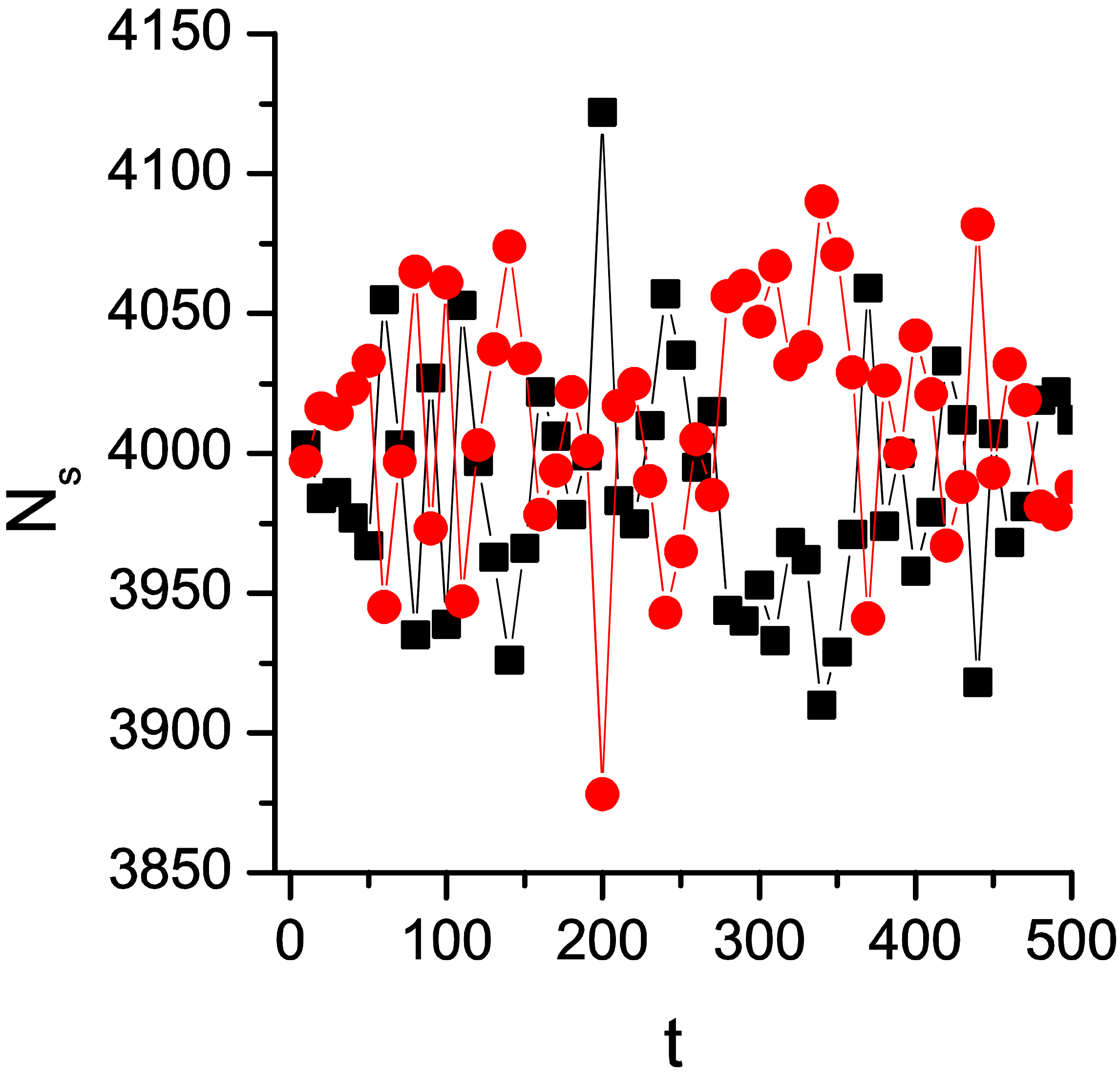

Figure 3

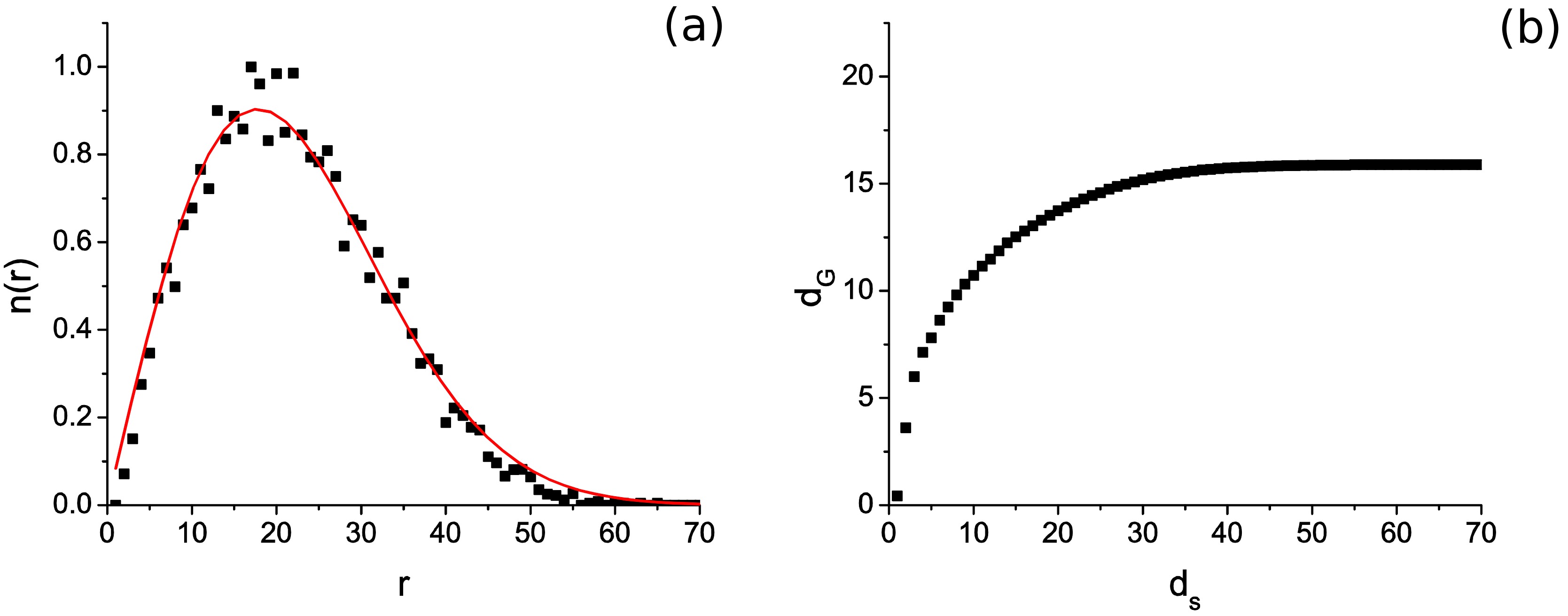

Figure 4

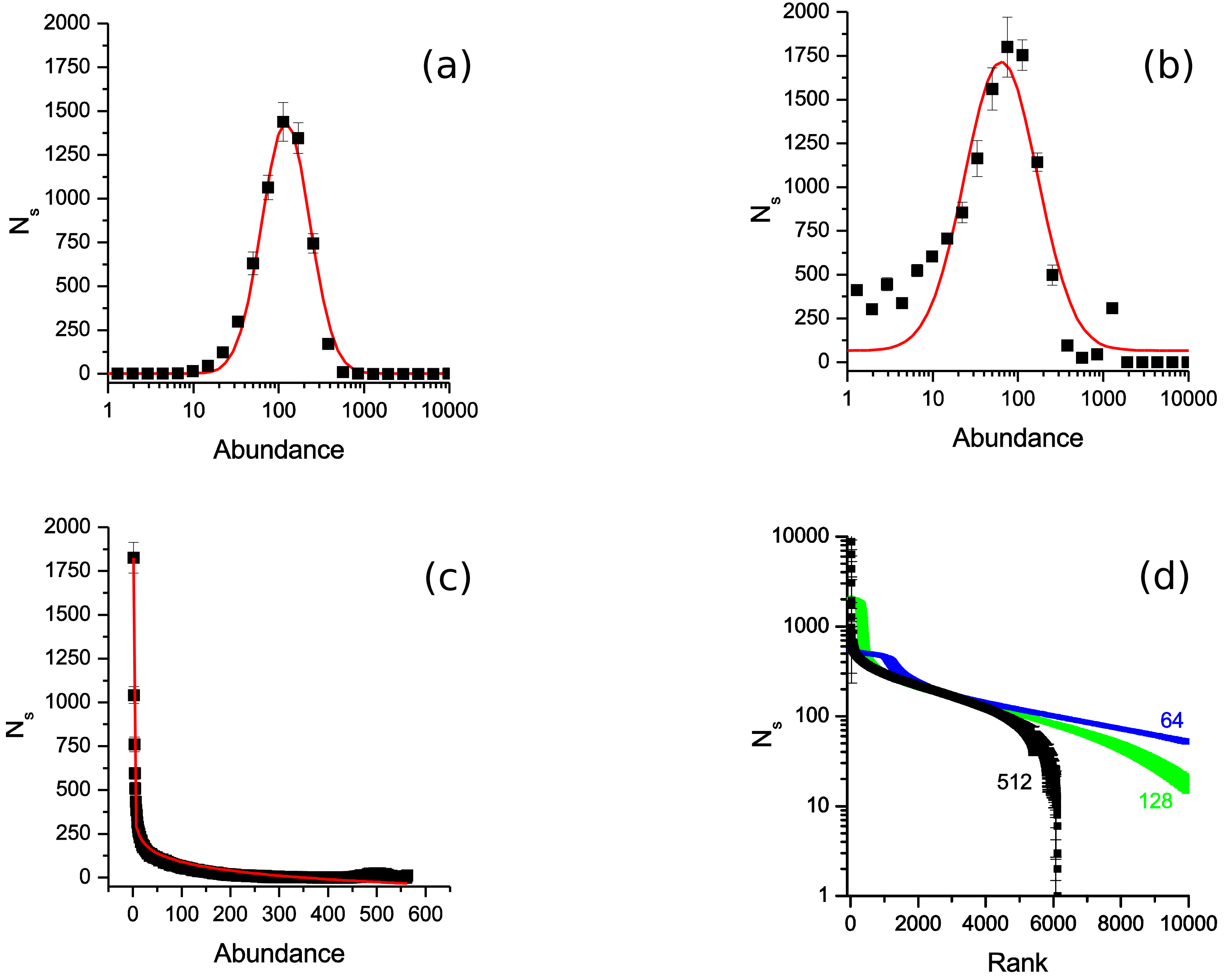

Figure 5

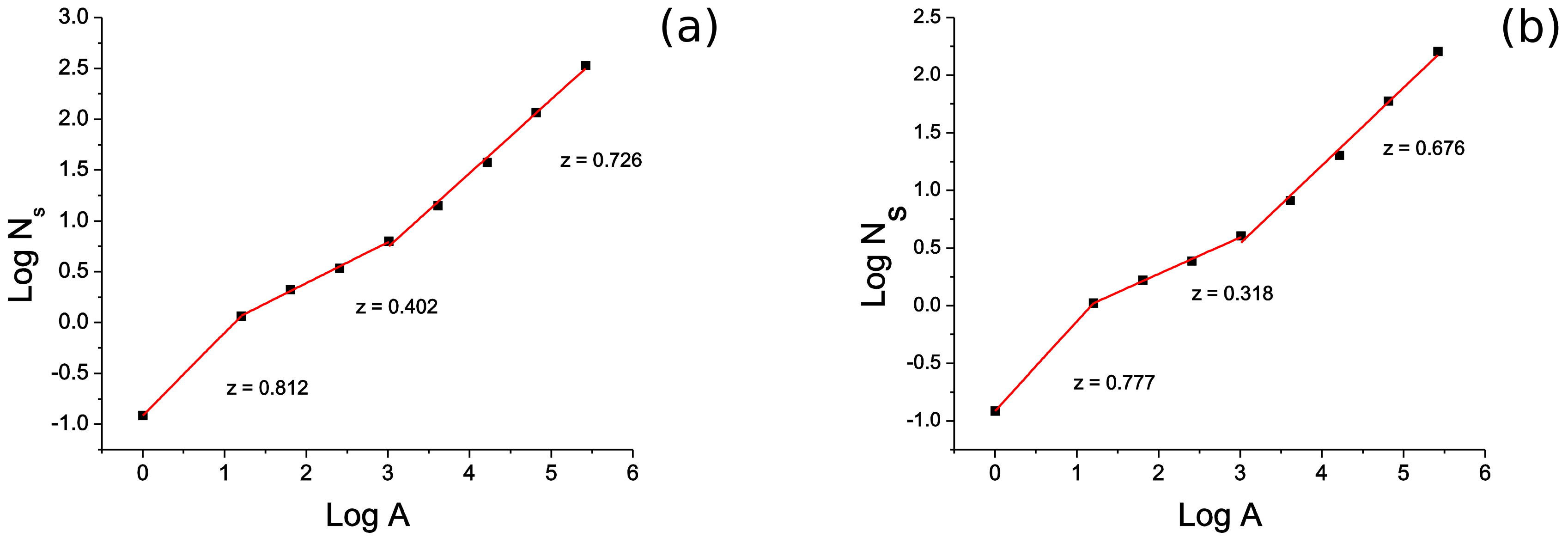

Figure 6

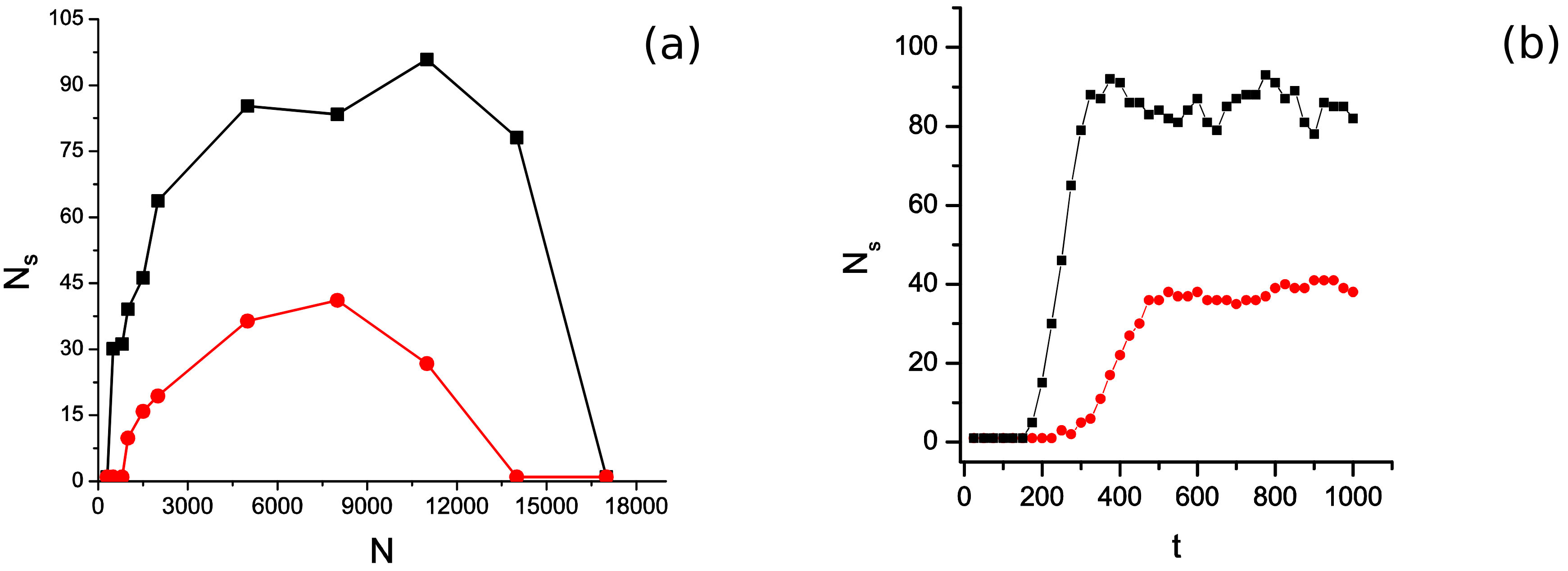

Figure 7

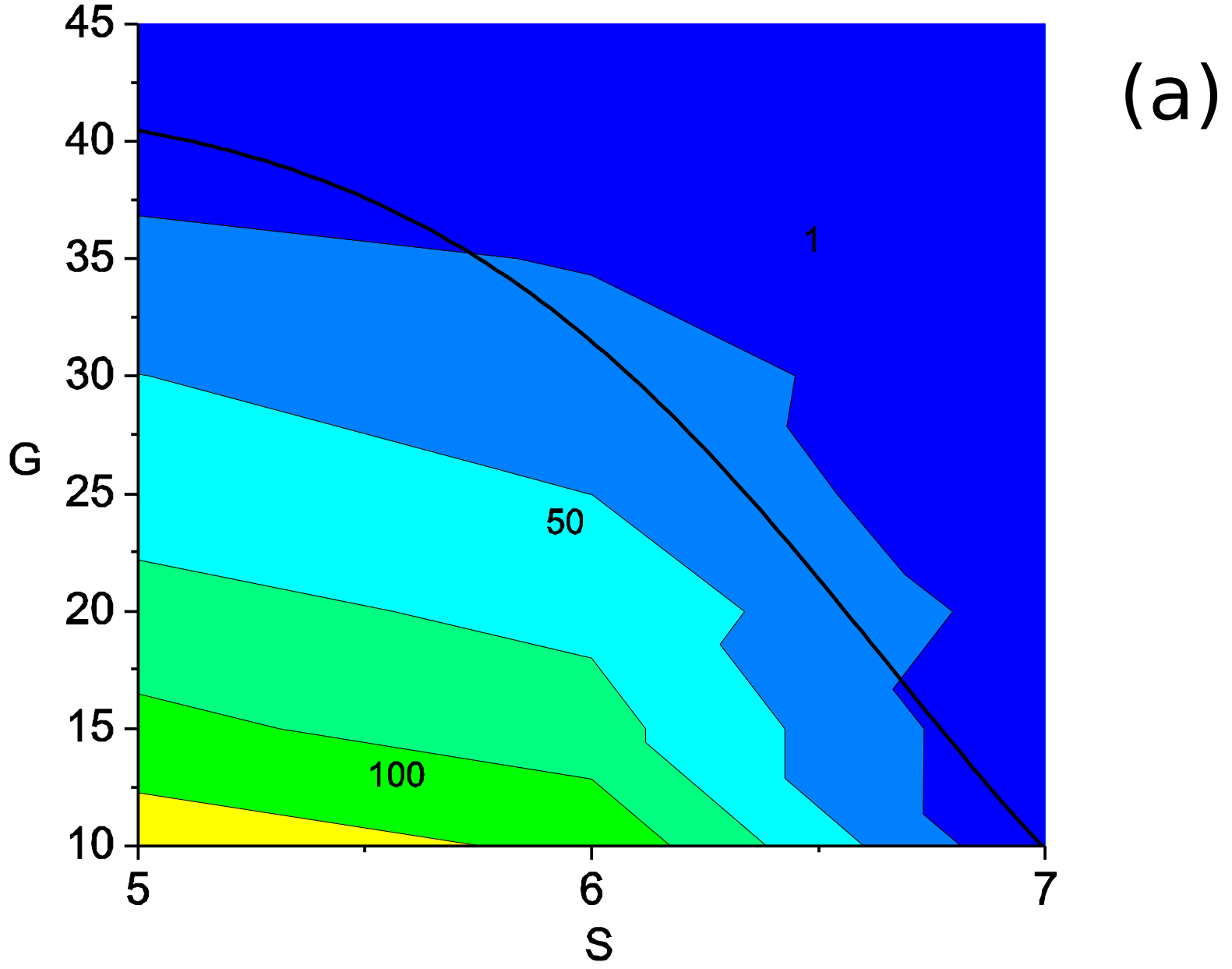

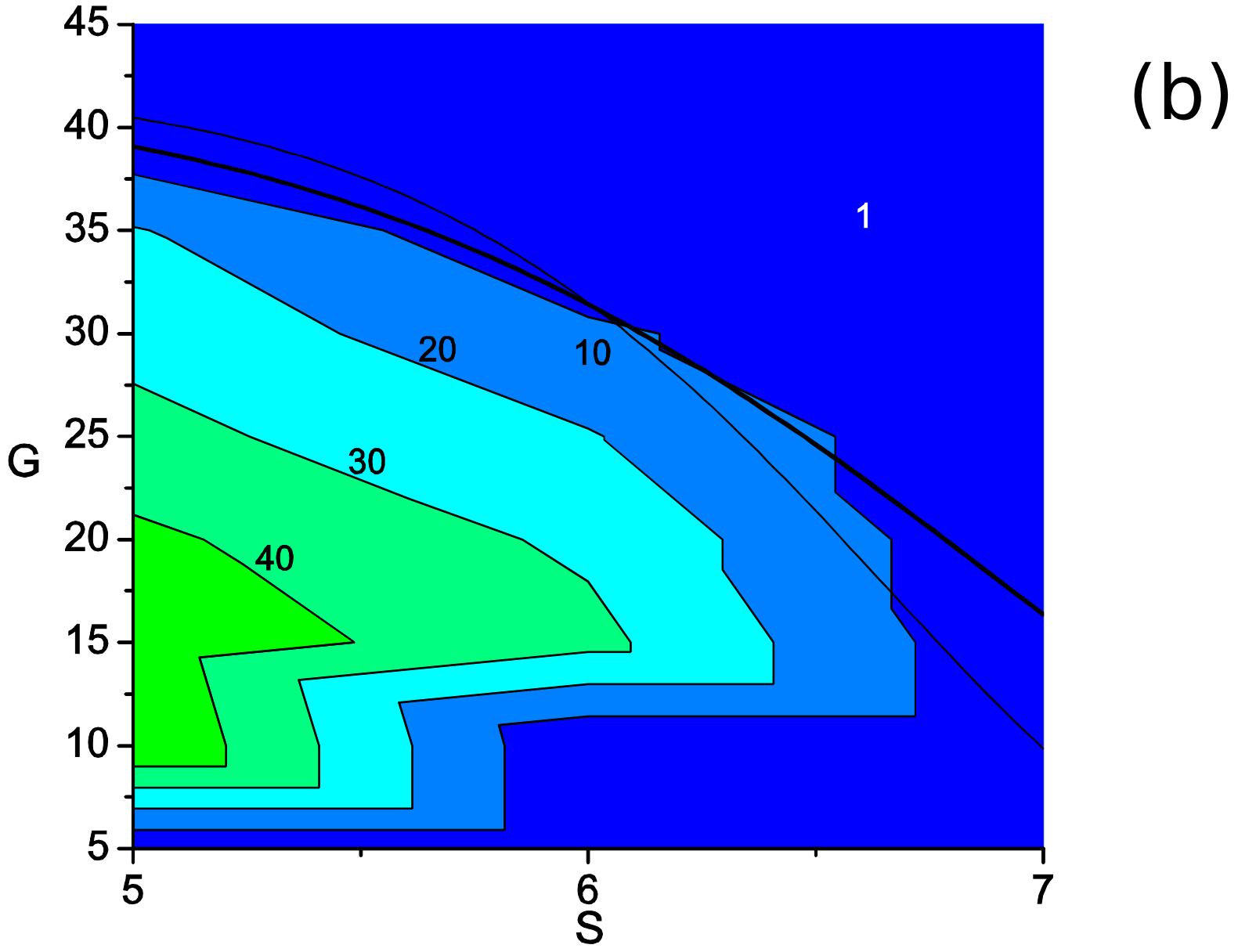

Figure 8